\documentstyle[12pt]{article}
 1
\font\twelverm=cmr10 scaled\magstep 1
 1

\font\tenbf=cmbx10
\font\tenrm=cmr10
\font\tenit=cmti10

\font\ninerm=cmr9

\font\eightrm=cmr8

\font\teni=cmmi10   \font\seveni=cmmi7  \font\fivei=cmmi5
\font\tensy=cmsy10
\font\sevenbf=cmbx7
\font\fivebf=cmbx5
\font\sevenrm=cmr7

\font\sevensy=cmsy7
\font\fiverm=cmr5

\font\fivesy=cmsy5

\def\tenpoint{\def\rm{\fam0\tenrm}
\textfont0=\tenrm \scriptfont0=\sevenrm \scriptscriptfont0=\fiverm
\textfont1=\teni \scriptfont1=\seveni \scriptscriptfont1=\fivei
\textfont2=\tensy \scriptfont2=\sevensy \scriptscriptfont2=\fivesy
\textfont3=\tenex \scriptfont3=\tenex \scriptscriptfont3=\tenex
\textfont\itfam=\tenit \def\it{\fam\itfam\tenit}
\textfont\slfam=\tensl \def\sl{\fam\slfam\tensl}
\textfont\ttfam=\tentt \def\tt{\fam\ttfam\tentt}
\textfont\bffam=\tenbf \scriptfont\bffam=\sevenbf
\scriptscriptfont\bffam=\fivebf \def\bf{\fam\bffam\tenbf}
\normalbaselineskip=12pt
\setbox\strutbox=\hbox{\vrule height8.5pt depth3.5pt width0pt}
\let\sc=\eightrm \let\big=\tenbig \normalbaselines\rm
}

\catcode`\@=11
\long\def\@makefntext#1{ 
\protect\noindent \hbox to 3.2pt {\hskip-.9pt
$^{{\ninerm\@thefnmark}}$\hfil}#1\hfill} 

\def\thefootnote{\fnsymbol{footnote}}
 \def\@makefnmark{\hbox to 0pt{$^{\@thefnmark}$\hss}}  

\def\ps@myheadings{\let\@mkboth\@gobbletwo
\def\@oddhead{\hbox{} 
\rightmark\hfil\ninerm\thepage}
\def\@oddfoot{}\def\@evenhead{\ninerm\thepage\hfil 
\leftmark\hbox{}}\def\@evenfoot{}
\def\sectionmark##1{}\def\subsectionmark##1{}}

\begin{document}
\newcommand \beq{\begin{eqnarray}}
\newcommand \eeq{\end{eqnarray}}
\newcommand \ga{\raisebox{-.5ex}{$\stackrel{>}{\sim}$}}
\newcommand \la{\raisebox{-.5ex}{$\stackrel{<}{\sim}$}}

\newcommand{\symbolfootnote}{\renewcommand{\thefootnote}
	{\fnsymbol{footnote}}}
\renewcommand{\thefootnote}{\fnsymbol{footnote}}
\newcommand{\alphfootnote}
	{\setcounter{footnote}{0}
	 \renewcommand{\thefootnote}{\sevenrm\alph{footnote}}}

\newcounter{sectionc}\newcounter{subsectionc}\newcounter{subsubsectionc}
\renewcommand{\section}[1] {\vspace{0.6cm}\addtocounter{sectionc}{1}
\setcounter{subsectionc}{0}\setcounter{subsubsectionc}{0}\noindent
	{\bf\thesectionc. #1}\par\vspace{0.4cm}}
\renewcommand{\subsection}[1] {\vspace{0.6cm}\addtocounter{subsectionc}{1}
	\setcounter{subsubsectionc}{0}\noindent
	{\it\thesectionc.\thesubsectionc. #1}\par\vspace{0.4cm}}
\renewcommand{\subsubsection}[1]
{\vspace{0.6cm}\addtocounter{subsubsectionc}{1}
	\noindent {\rm\thesectionc.\thesubsectionc.\thesubsubsectionc.
	#1}\par\vspace{0.4cm}}
\newcommand{\nonumsection}[1] {\vspace{0.6cm}\noindent{\bf #1}
	\par\vspace{0.4cm}}

\newcounter{appendixc}
\newcounter{subappendixc}[appendixc]
\newcounter{subsubappendixc}[subappendixc]
\renewcommand{\thesubappendixc}{\Alph{appendixc}.\arabic{subappendixc}}
\renewcommand{\thesubsubappendixc}
	{\Alph{appendixc}.\arabic{subappendixc}.\arabic{subsubappendixc}}

\renewcommand{\appendix}[1] {\vspace{0.6cm}
        \refstepcounter{appendixc}
        \setcounter{figure}{0}
        \setcounter{table}{0}
        \setcounter{equation}{0}
        \renewcommand{\thefigure}{\Alph{appendixc}.\arabic{figure}}
        \renewcommand{\thetable}{\Alph{appendixc}.\arabic{table}}
        \renewcommand{\theappendixc}{\Alph{appendixc}}
        \renewcommand{\theequation}{\Alph{appendixc}.\arabic{equation}}
        \noindent{\bf Appendix \theappendixc #1}\par\vspace{0.4cm}}
\newcommand{\subappendix}[1] {\vspace{0.6cm}
        \refstepcounter{subappendixc}
        \noindent{\bf Appendix \thesubappendixc. #1}\par\vspace{0.4cm}}
\newcommand{\subsubappendix}[1] {\vspace{0.6cm}
        \refstepcounter{subsubappendixc}
        \noindent{\it Appendix \thesubsubappendixc. #1}
	\par\vspace{0.4cm}}

\def\abstracts#1{{
	\centering{\begin{minipage}{30pc}\tenpoint\baselineskip=12pt\noindent
	\centerline{\tenpoint ABSTRACT}\vspace{0.3cm}
	\parindent=0pt #1
	\end{minipage} }\par}}

\newcommand{\bibit}{\it}
\newcommand{\bibbf}{\bf}
\renewenvironment{thebibliography}[1]
	{\begin{list}{\arabic{enumi}.}
	{\usecounter{enumi}\setlength{\parsep}{0pt}
\setlength{\leftmargin 1.25cm}{\rightmargin 0pt}
	 \setlength{\itemsep}{0pt} \settowidth
	{\labelwidth}{#1.}\sloppy}}{\end{list}}

\topsep=0in\parsep=0in\itemsep=0in
\parindent=1.5pc

\newcounter{itemlistc}
\newcounter{romanlistc}
\newcounter{alphlistc}
\newcounter{arabiclistc}
\newenvironment{itemlist}
    	{\setcounter{itemlistc}{0}
	 \begin{list}{$\bullet$}
	{\usecounter{itemlistc}
	 \setlength{\parsep}{0pt}
	 \setlength{\itemsep}{0pt}}}{\end{list}}

\newenvironment{romanlist}
	{\setcounter{romanlistc}{0}
	 \begin{list}{$($\roman{romanlistc}$)$}
	{\usecounter{romanlistc}
	 \setlength{\parsep}{0pt}
	 \setlength{\itemsep}{0pt}}}{\end{list}}

\newenvironment{alphlist}
	{\setcounter{alphlistc}{0}
	 \begin{list}{$($\alph{alphlistc}$)$}
	{\usecounter{alphlistc}
	 \setlength{\parsep}{0pt}
	 \setlength{\itemsep}{0pt}}}{\end{list}}

\newenvironment{arabiclist}
	{\setcounter{arabiclistc}{0}
	 \begin{list}{\arabic{arabiclistc}}
	{\usecounter{arabiclistc}
	 \setlength{\parsep}{0pt}
	 \setlength{\itemsep}{0pt}}}{\end{list}}

\newcommand{\fcaption}[1]{
        \refstepcounter{figure}
        \setbox\@tempboxa = \hbox{\tenpoint Fig.~\thefigure. #1}
        \ifdim \wd\@tempboxa > 6in
           {\begin{center}
        \parbox{6in}{\tenpoint\baselineskip=12pt Fig.~\thefigure. #1 }
            \end{center}}
        \else
             {\begin{center}
             {\tenpoint Fig.~\thefigure. #1}
              \end{center}}
        \fi}

\newcommand{\tcaption}[1]{
        \refstepcounter{table}
        \setbox\@tempboxa = \hbox{\tenpoint Table~\thetable. #1}
        \ifdim \wd\@tempboxa > 6in
           {\begin{center}
        \parbox{6in}{\tenpoint\baselineskip=12pt Table~\thetable. #1 }
            \end{center}}
        \else
             {\begin{center}
             {\tenpoint Table~\thetable. #1}
              \end{center}}
        \fi}

\def\@citex[#1]#2{\if@filesw\immediate\write\@auxout
	{\string\citation{#2}}\fi
\def\@citea{}\@cite{\@for\@citeb:=#2\do
	{\@citea\def\@citea{,}\@ifundefined
	{b@\@citeb}{{\bf ?}\@warning
	{Citation `\@citeb' on page \thepage \space undefined}}
	{\csname b@\@citeb\endcsname}}}{#1}}

\newif\if@cghi
\def\cite{\@cghitrue\@ifnextchar [{\@tempswatrue
	\@citex}{\@tempswafalse\@citex[]}}
\def\citelow{\@cghifalse\@ifnextchar [{\@tempswatrue
	\@citex}{\@tempswafalse\@citex[]}}
\def\@cite#1#2{{$\null^{#1}$\if@tempswa\typeout
	{IJCGA warning: optional citation argument
	ignored: `#2'} \fi}}
\newcommand{\citeup}{\cite}

\def\fnm#1{$^{\mbox{\scriptsize #1}}$}
\def\fnt#1#2{\footnotetext{\kern-.3em
	{$^{\mbox{\sevenrm #1}}$}{#2}}}


\centerline{\tenbf QUARK MATTER DROPLET FORMATION IN NEUTRON STARS}
\vspace{0.8cm}
\centerline{\tenrm H. HEISELBERG}
\baselineskip=13pt
\centerline{\tenit NORDITA, Blegdamsvej 17, DK-2100 Copenhagen \O, Denmark}
\vspace{0.9cm}
\abstracts{
The formation rate of quark matter droplets in neutron stars is
calculated from a combination of bubble formation rates in
cold degenerate and high temperature matter. Nuclear matter
calculations of the viscosity and thermal conductivity are applied.
Results show that droplets form only in the core of neutron stars
shortly after supernova collapse, where pressures and temperatures
are high, and for sufficiently small interface tension between nuclear
and quark matter. Coulomb energies hinder formation of large droplets whereas
the presence of strange hadrons in nuclear matter increase the droplet
formation rate.
}
\vglue 0.3cm
\twelverm   
\baselineskip=14pt

\section{Introduction}
Quark matter in neutron star has recently been found to be able to coexist with
nuclear matter and have a rich structure in this mixed phase\cite{G,HPS}.
This is in
contrast to standard models\cite{ALL} where neutron stars have a pure quark
matter core with a mantle of nuclear matter outside. The quark and nuclear
matter mixed phase appear at much lower densities. A lattice of quark matter
droplets appear already around twice nuclear matter saturation densities
which increases the probability for the existence of quark matter in neutron
stars.

We shall here discuss the formation of quark matter droplets and calculate the
rate by which the first quark matter droplets form at the high densities in
centers of neutron stars. (If quark matter is the true
ground state of hadronic matter in vacuum\cite{Witten}, such
strangelets would probably reside inside the stars as relics from the early
universe and would convert the whole star to a strange star.) We
combine bubble formation rates of both cold degenerate and hot systems
and apply transport coefficients of nuclear matter.
 To estimate the time scales for the droplet formation the pressure and
temperatures are calculated from standard equation of states used in
supernova core
collapse for nuclear matter and a simple Bag model for quark matter.

\section{Droplet Free Energy}

In the core of a supernova collapse or the later neutron star the nuclear
matter may be dense enough that it is energetically favorable to form a quark
matter droplet. The work forming a droplet of radius $R$ is
\beq
   W &=& -\frac{4\pi}{3}R^3 \Delta P + 4\pi\sigma R^2
      + N_q\Delta\mu + E_C   , \label{F}
\eeq
where $\Delta P=P_{QM}-P_{NM}$ is the pressure difference, $\sigma$ the surface
tension, $\Delta\mu=\mu_{QM}-\mu_{NM}$
the difference in chemical potentials between nuclear and quark matter,
$N_q=\frac{4\pi}{3}R^3n_{QM}$ the number of quarks in the droplet and $E_C$ the
Coulomb energy of the droplet.

It is usually assumed that the embedded droplet is in chemical equilibrium with
its surroundings at nucleation\cite{LL}, i.e.,
\beq
    \mu_n = \mu_u+2\mu_d \, , \quad\quad
    \mu_p = 2\mu_u+\mu_d \, . \label{mu}
\eeq
Olesen and Madsen\cite{OM} consider
the instantaneous collapse of the nuclear matter into a quark matter
containing the original flavor composition.
The resulting droplet will be out of chemical equilibrium
and diffusion of various particles will take place until chemical equilibrium
is attained.
We shall take the standard approach that the quark matter is in chemical
equilibrium with its surroundings since we expect the diffusion of nucleons
in and out of the droplet to be at least as fast as the formation of a large
droplet. Furthermore, we shall assume the presence of sufficient
strangeness in nuclear
matter from strange hadrons like $\Sigma^-$, $K^-$, $\Lambda$... . With
sufficient rapid diffusion of strangenes during droplet formation we can then
also assume $\mu_d=\mu_s$ inside the quark matter droplet. Otherwise, only
{\it u, d} quark matter droplets can form since strangeness production only
occur slowly on weak decay time scales. Pure {\it u,d}- quark matter will
have higher free energies and are thus harder to form.

Let us first discuss small droplets where Coulomb effects can be
neglected. Since the maximal temperatures reached in supernovae are much less
than the chemical potentials $T\ll\bar{\mu}$ we can also ignore thermal
energies in (\ref{F}).  The nucleation rate contains the usual Boltzmann factor
in which the work to form a critical bubble enters. The critical bubble is
determined by the maximum of Eq. (\ref{F}) which gives a critical bubble
radius,
\beq
   R_c = \frac{2\sigma}{\Delta P} \, , \label{Rc}
\eeq
at which the work is
\beq
   W_c = \frac{16\pi}{3}\frac{\sigma^3}{\Delta P^2} \, . \label{Fc}
\eeq
The pressure difference across the boundary of the critical bubble
is balanced by the surface tension, $P_{QM}=P_{NM}+2\sigma/R_c$.
The temperatures and chemical potentials are the same in the two phases and
therefore  the critical droplet is in complete
phase equilibrium with the nuclear matter when created.
The surface tension is unfortunately a poorly determined parameter and
we refer to Ref.\cite{HPS} for a discussion. It is probably within the
range $\sigma\sim 10-200$MeV/fm$^2$.

The Coulomb energy, $(3/5)Z^2e^2/R$, grows rapidly with size for
a droplet with constant charge density\cite{HPS}.
Screening or charge rearrangement
reduces the Coulomb energy significantly \cite{scr} but still the critical
radius and work are found to increase due to Coulomb effects. A more
detailed analysis shows that the formation of droplets more
than a few {\it fm} in size is severely hindered.

\section{Formation Rate}
The droplet formation rate was calculated by Langer \& Turski\cite{LT}
\beq
    I = \frac{\kappa}{2\pi} \Omega_0 \, e^{-W_c/T} \, . \label{I}
\eeq
The  ``statistical" prefactor
\beq
    \Omega_0 = \frac{2}{3\sqrt{3}} \left(\frac{\sigma}{T}\right)^{3/2}
               \left(\frac{R_c}{\xi_q}\right)^4 \, , \label{Omega}
\eeq
measures the phase-space volume of the saddle point around $R_c$
that the droplet has to pass on its way to the lower energy state.
Here $\xi_q$ is the quark correlation length which also estimates the surface
thickness. We shall assume that it is of order the QCD Debye screening
length $\xi_q\simeq 1/q_D$ where
$ q_D^2 = (2/\pi) \alpha_s \sum_q \mu_q^2
   = (2/\pi) \alpha_s N_q \bar{\mu}^2 $.
Here, $\alpha_s$ is the QCD fine-structure constant and $N_q$ the number of
quark flavors present. With $\alpha_s\simeq 0.3$, $N_q=3$ and
$\bar{\mu}\simeq 400$ MeV we have $\xi_q=1/q_D=0.7$ fm.
Note that Eq. (\ref{Omega}) is only valid when $R_c\ga\xi_q$ which
is the case for reasonable estimates of $\sigma$ and $\Delta P$.
 The ``dynamical" prefactor determines the droplet growth rate and is
\beq
    \kappa =  \frac{2\sigma}{\Delta w^2 R_c^3}
       \left(\lambda T + 2(\frac{4}{3}\eta+\zeta)\right) \, ,\label{kappa}
\eeq
where $\Delta w=w_{QM}-w_{NM}$ is the enthalpy difference and
$\lambda$, $\eta$ and $\zeta$ are the thermal conductivity, shear and
bulk viscosities respectively. The dynamical prefactor (\ref{kappa})
incorporates both the low temperature result of Langer \& Turski,
where the viscous term is ignored, and the high temperature result of
Csernai \& Kapusta\cite{CK}, where the thermal conductivity is negligible.
Since both calculations were performed in the hydrodynamic limit linearized
in the transport coefficients, the two results simply add up\cite{VV}.

The shear and bulk viscosities and the thermal conductivity have been
estimated by Danielewicz\cite{Dan} within the Boltzmann equation
with an effective nucleon-nucleon differential scattering cross section.
The contributions to the dynamical prefactor from the bulk viscosity
and the thermal conductivity are negligible as compared to that from the
viscosity which typically is $\eta\sim 50$MeV/fm$^2$.
The result of Langer and Turski\cite{LT} include only the thermal
conduction term and
is therefore not applicable to the case of degenerate
quark matter droplet formation.
For the formation of hadronic bubbles in quark matter one can use the
perturbative QCD calcuations of transport coefficients\cite{deg}.
Here the viscous damping dominates as well and the shear viscosity is
even larger than in nuclear matter.

Still neglecting Coulomb energies we obtain from
Eqs. (\ref{Fc},\ref{I},\ref{Omega})
\beq
    I &=& \frac{\sigma_{20}^{7/2}\, \bar{\mu}_{400}^2\, \eta_{50} }
         {\Delta P_{10} \Delta w_{10}^2\,  T_{10}^{3/2} }
     \exp\left[ 185-134\frac{\sigma_{20}^3}{\Delta P_{10}^2 T_{10} }
      \right]   {\rm s}^{-1} {\rm km}^{-3} \, , \label{In}
\eeq
assuming $\alpha_s=0.3$ and $N_q=3$. Here
we have written all the quantities in typical units
\cite{HPS,Jan}: the
enthalpy and pressure differences
in units of 10 MeV/fm$^3$, the quark chemical potential in 400 MeV,
the temperature in 10 MeV, the surface tension in units of
20 MeV/fm$^2$ and the viscosity in 50 MeV/fm$^2$.
The resulting rate is very sensitive to $\sigma$, $\Delta P$ and $T$ .
It is rather insensitive to the prefactor and therefore also the
viscosity. Using the prefactor $T^4$ in stead reduces the rate by approximately
11 orders of magnitude which does, however, not have much effect in
(\ref{In}). Here the huge exponent 185 arise mainly from conversion
of hadronic scales (fm$^{-4}$) to neutron star dimensions and cooling times
(km$^{-3}$s$^{-1}$).

The pressure difference depends on the equation of state.  It increases
with density and so droplet formation is most likely in the center of the
neutron star.  The maximum temperature achieved in supernova explosions is
largest in the core with a value around 10 MeV (depending on the equation of
state and details of the supernova collapse) and it cools down to a few MeV in
10-20 seconds\cite{Jan}.
Consequently, the droplets predominantly form in the neutron star
center within few seconds after the core collapse when temperatures are
highest as already pointed out in (\cite{OM,Alcock,Hor}) and only if the
surface
tension is sufficiently low.

The total number of droplets formed is the integrated rate
over volume of the neutron star and time after the
supernova core collapse
\beq
   N = \int_0^{R_{NS}} 4\pi r^2dr\int_{t_{SN}}^\infty dt
               \,   I(\Delta P(r),T(t)) \, . \label{N}
\eeq
The rate depends mainly on the spatial coordinate through the
pressure difference and on time through the temperature.
The pressure and temperature depends sensitively on the equation of state
of nuclear and quark matter and on the details of the supernova collapse.

To convert the core of the neutron star into quark matter, whether it is
in a mixed phase with nuclear matter or not, we must form at least one
droplet, i.e. $N>1$. According to (\ref{In}) and (\ref{N}) this requires
\beq
    \sigma \, \la\,  24 \, {\rm MeV}\cdot{\rm fm}^{-2} \,  \cdot
         \, \Delta P_{c,10}^{2/3}\,  T_{c,10}^{1/3} \quad , \label{s}
\eeq
where $P_c$ and $T_c$ are now the core values. This condition is
almost insensitive to the size and cooling time of the hot core.

\section{Summary}
A formation rate for quark matter droplets has been derived
applying transport coefficients from nuclear matter calculations. The resulting
rate is very sensitive to the surface tension at the nuclear/quark matter
interface and to the pressure difference and temperature.
For large droplets it is important to include Coulomb enegies which reduce
formation rate for droplets larger than a few $\sim fm$
drastically. The presence of strange hadrons and sufficiently fast
strange quark diffusion allow the formation of strange quark matter
droplets which requires less work than purely {\it u,d}-quark droplets
and thus improve the production rate.
Quark matter
droplets form only in the core of neutron stars, where the pressure
difference is highest, and shortly after the core collapse in supernovae
explosions, where temperatures are highest. Eq. (\ref{s}) gives the
necessary conditions for creating quark matter in neutron stars.
The pressure difference depends strongly on the equation of state used
for both nuclear and quark matter and with the present uncertainties in
both and in the surface tension, the droplet formation rate cannot be reliably
estimated.
If neutron stars burn into quark stars\cite{deg} it would
happen right after the supernova
explosion, and so this mechanism is unlikely to be the cause of gamma ray
bursters.

\section{Acknowledgements}
This work was supported by DOE grant DE-AC03-76SF00098
and the Danish Natural Science Research Council.

\end{document}